\documentclass[12pt]{article}
\usepackage{epsfig,palatino,euler}

\usepackage{latexsym}
\usepackage{float, dcolumn}
\usepackage{graphicx}
\usepackage{amsfonts, amsbsy}
\usepackage{amsmath, subfigure}
\usepackage{amscd}
\usepackage[round,colon]{natbib}
\bibpunct[,]{(}{)}{;}{a}{ }{,}

\newlength{\pwidth}
\newlength{\fwidth}
\begin{document}

\vfill

\title{
Lagrangian transport through an ocean front in the North-Western
Mediterranean Sea}

\vfill

\author{\Large Ana M. Mancho \\
Instituto de Matem{\'a}ticas y F{\'\i}sica Fundamental,\\
Consejo Superior de Investigaciones Cient{\'\i}ficas (CSIC), \\
Serrano 121, 28006 Madrid, Spain \\
\Large Emilio Hern{\'a}ndez-Garc{\'\i}a \\
Instituto Mediterr{\'a}neo de Estudios Avanzados, \\
IMEDEA (CSIC - Universitat de les Illes Balears) \\
E-07122 Palma de Mallorca, Spain \\
\Large Des Small and Stephen Wiggins \\
School of Mathematics, University of Bristol, \\
Bristol BS8 1TW, United Kingdom \\
\Large Vicente Fern{\'a}ndez \\
Istituto Nazionale di Geofisica e Vulcanologia INGV, \\
Via Donato Creti 12, 40128 Bologna, Italy }

\vfill

\date{June 29, 2006}

\maketitle

\newpage

\begin{abstract}

We analyze with the tools of lobe dynamics the velocity field from
a numerical simulation of the surface circulation in the
Northwestern Mediterranean Sea. We identify relevant hyperbolic
trajectories and their manifolds, and show that the transport
mechanism known as the `turnstile', previously identified in
abstract dynamical systems and simplified model flows, is also at
work in this complex and rather realistic ocean flow. In addition
nonlinear dynamics techniques are shown to be powerful enough to
identify the key geometric structures in this part of the
Mediterranean. In particular the North Balearic Front, the
westernmost part of the transition zone between saltier and
fresher waters in the Western Mediterranean is interpreted in
terms of the presence of a semipermanent ``Lagrangian barrier''
across which little transport occurs. Our construction also
reveals the routes along which this transport happens. Topological
changes in that picture, associated with the crossing by eddies
and that may be interpreted as the breakdown of the front, are
also observed during the simulation.

\end{abstract}

\newpage

\section{Introduction}
\label{sec:intro}

Ocean water masses of different origins have distinct contents of
salt, heat, nutrients and chemicals. Currents transport them, and
energetic mesoscale features are responsible for most of their
mixing with surrounding waters.

Vortices are the most well studied of such structures. Frequently
they are long lived, and their cores remain relatively protected
from the neighboring areas, so that water trapped inside could
maintain its biogeochemical properties for long time, being
transported with the vortex. In steady horizontal velocity fields,
the centers of vortices are readily identified as elliptic points
in the streamfunction level sets. The presence of closed
streamlines around them is the mathematical reason for the
isolation of the vertex core from the exterior fluid.

When the velocity field changes in time, closed streamlines are
replaced by more complex structures, some of which can be related
in idealized cases to the Kolmogorov-Arnold-Moser tori of
dynamical systems theory. For slowly varying velocity fields
vortex cores remain coherent during some time, but there is
vigourous stirring of the surrounding fluid that finally leads to
water mixing. In order to understand this mixing process, one
should focus on features of the velocity field different from the
elliptic points characterizing the vortex cores. Since some time
ago \citep{jsw,idw,msw2} hyperbolic trajectories (trajectories
with saddle-like stability properties which are solutions to a
dynamical system) have been recognized as the structures
responsible for most of the stretching and generation of
intertwined small scales that finally leads to mixing. In
particular there are {\sl Distinguished Hyperbolic Trajectories}
(DHTs), characterized by its special persistence as compared with
the other hyperbolic structures, that act as the organizing
centers of the fluid stirring processes.

Despite the great amount of attention devoted to the
identification and characterization of vortices and their dynamics
in oceanographic contexts
\citep{Olson1991,Puillat2002,Ruiz2002,Isern2006}, there are few
studies focussed on hyperbolic objects, and most of them in
idealized settings. A possible reason for this may be that the
intrinsic instability of trajectories close to hyperbolic points
makes more difficult their identification, in contrast with the
recurrent character of trajectories in vortices that allow their
tracking for long times from in situ and satellite measurements.
In addition, whereas many aspects of vortex dynamics can be
analyzed in a Eulerian framework, hyperbolic trajectories are
defined by their Lagrangian characteristics and only in this frame
can they be fully characterized \citep{idw, msw2}.

In this Paper we identify relevant hyperbolic trajectories from
the surface velocity field  of the Western Mediterranean Sea,
obtained from a three dimensional model simulation under
climatological atmospheric forcing. Our aim is to show that
transport mechanisms, in particular the so called `turnstile'
mechanism, previously identified in abstract dynamical systems
\citep{channon, bartlett, mmp2, w3}, and discussed in the context
of rather simple model flows \citep{rlw, blw, sam, dw}, are also
at work in this complex and rather realistic ocean flow. More
broadly, nonlinear dynamics techniques are shown to be powerful
enough to identify the key geometric structures in this part of
the Mediterranean.

Western Mediterranean surface layers (up to a depth of about 150
m) contain Modified Atlantic Waters of rather different
characteristics \citep{Millot1999}. Fresher waters (salinity about
36.5 psu) recently entered from the Atlantic occupy the Algerian
Basin at the South and older and saltier waters (salinity above 38
psu) occupy the Northern part of the area. The dynamics of the
contact zone between these two surface water masses, and
eventually their mixing, is important to understand the physical
and biogeochemical properties of the Western Mediterranean. We
focus on one of the oceanographic structures known to be of
importance in these processes, the so-called North Balearic Front
\citep{LopezGarcia1994,Pinot95,Millot1999}. It extends roughly
along the Southwest-Northeast direction at the North of the
Balearic Islands, with significant displacements and deformations.
It is characterized by a strong salinity jump of about 0.6 psu
(37.4 psu to the South and 38.0 to the North) down to 150 m depth.
This identifies the front as the main transition zone between the
two water masses. In Winter a weak but detectable temperature
gradient of 0.5-1 K/5 km  can be observed in satellite images.

After selecting an interval of time in our simulation during which
the front is well formed, we explore the transport properties of
the surface velocity field in the region, and find the relevant
Lagrangian structures, hyperbolic points and their manifolds,
responsible for the establishment and permanence of the front. The
location of the front is identified as a ``Lagrangian barrier''
across which transport is small (as quantified with the tools of
lobe dynamics), and occurs via filaments that entrain water along
transport routes that we identify. The presence of eddies strongly
affect the Lagrangian structures in a process that can be
interpreted as the break-down of the front.

The Paper is organized as follows: In Section
\ref{sec:trajectories} we introduce some of the dynamical systems
concepts that will be used in the following. Section
\ref{sec:model} describes our numerical ocean model and Section
\ref{sec:computation} addresses the adaptation of the standard
algorithms to the kind of oceanographic data provided by the
model. Section \ref{sec:results} contain our main results and the
conclusions are summarized in Section \ref{sec:conclusions}.

\section{Distinguished Hyperbolic Trajectories and their Manifolds}
\label{sec:trajectories}

In recent years there have been many applications of the dynamical
systems approach to transport in oceanographic flows.  Recent
reviews are \citet{jw2, warfm, samwig}. In this section we
describe the basic ideas that we will use in our analysis.
Although the same concepts are of use in three dimensional flows,
we describe here just the two dimensional situation, since as
discussed below the structures we identify can be considered two
dimensional to a good approximation during the time scales
relevant here.

Stagnation points are well known features of steady flows that
generally play an important role in ``organizing'' the
qualitatively distinct streamlines in the flow. For example,
saddle-type stagnation points can occur on boundaries where
streams of flow tangential to the boundary coming from opposite
directions meet and then separate from that boundary. Saddle-type
stagnation points can occur in the interior of a flow at a point
where fluid seems to both converge to the point along two opposite
directions and diverge from it along two different directions. In
steady flows the stagnation point is a trivial example of a fluid
particle trajectory (i.e., it is a ``solution,'' in this case a
fixed point solution, of the equations for fluid particle motions
defined by the velocity field) and the saddle point nature is
manifested by the fact that there are directions for which nearby
trajectories approach the stagnation point at an exponential rate
and move away from the stagnation point at an exponential rate.
These directions are sometimes referred to as ``stagnation
streamlines''. They define material curves that ``cross'' at the
stagnation point and typically form the boundaries between
qualitatively distinct regions of flow.

A related ``time-dependent'' picture as that described above
exists for unsteady flows, with both similar, as well as much more
complex, implications for transport. In unsteady flows a
stagnation point {\sl at a given time} --or rather, an {\sl
Instantaneous Stagnation Point} (ISP)-- is a location  at which
velocity vanishes {\sl at that time}. The sequence of locations is
generally {\sl not} a fluid particle trajectory \citep{idw}. The
true analog of the saddle-type stagnation point of steady flows is
a {\em Distinguished Hyperbolic Trajectory} (DHT) \citep{idw}.
``Hyperbolic'' is the dynamical systems terminology for
``saddle-type''. These are fluid particle trajectories that have
(time-dependent) directions for which nearby trajectories approach
and move away from the DHT at exponential rates. ``Distinguished''
is a notion that is discussed in detail in \citet{idw}, but the
idea is that these are the key, {\sl isolated} hyperbolic
trajectories that serve to organize the transport behavior in a
flow because they remain substantially more localized (in a well
defined sense, see \citet{idw}) than neighboring hyperbolic
trajectories. \citet{idw}, \citet{jsw}, and \citet{msw2} develop
the algorithms that allow us to compute DHTs in a given flow. They
are iterative methods that start with a first guess for the DHT
positions in an interval of time (i.e. an initial curve in space
and time) and then refine the space-time curve by imposing the
criteria of hyperbolicity and localization. In the flows
considered here, a good first guess is the location of ISPs, since
it turns out that a DHT is often found in the neighborhood of an
Eulerian ISP.

Just as in the steady case, there are analogs to the stagnation
streamlines: in the dynamical systems terminology these are
referred to as the {\em stable and unstable manifolds of the DHT},
and they are time-dependent material curves. In dynamical systems
terminology the fact that they are material curves means that they
are {\em invariant}, i.e. a fluid particle trajectory starting on
one of these curves must remain on that material curve during the
course of its time evolution. ``Stable manifold'' means that
trajectories starting on this material curve approach the DHT at
an exponential rate as time goes to infinity, and ``unstable
manifold'' means that trajectories starting on this material curve
approach the DHT at an exponential rate as time goes to minus
infinity. \citet{mswi,msw2} develop the algorithms that enable us
to compute the stable and unstable manifolds of hyperbolic
trajectories.

In unsteady flows stable and unstable manifolds of DHTs can
intersect in isolated points different from the DHTs. This is a
fundamental difference with respect to steady flows, and give rise
to moving regions of fluid bounded by pieces of stable and
unstable manifolds, the so called ``lobes''. Since the manifolds
are material lines, fluid can not cross them by purely advective
processes and thus they are perfect Lagrangian barriers
(diffusion, or motion along the third dimension can however induce
cross-manifold transport). Motion of the ``lobes'' is thus the
mechanism responsible for mediating Lagrangian transport between
different regions. References describing ``lobe dynamics'' in
general are \citet{rw, blw, w3, maw, samwig}. Examples of
applications of lobe dynamics to oceanographic flows are
\citet{ns, rmpj, ypj, ypj2, mphj, deese}. We will describe these
ideas more fully in the context of transport associated with the
Balearic front in the Mediterranean.

\section{The Ocean Circulation Model}
\label{sec:model}

In this work we analyze velocity fields which are obtained from an
ocean model, DieCAST \citep{Dietrich1997}, adapted to the
Mediterranean Sea \citep{Dietrich2004,Fernandez2005}. The 3D
primitive equations are discretized with a fourth-order collocated
control volume method. In zones adjacent to  boundaries a
conventional second order method is used. A fundamental feature of
control volume based models is that the predicted quantities are
control volume averages, while face-averaged quantities are used
to evaluate fluxes across control volume faces
\citep{Sanderson1998}. These quantities are computed using
fourth-order approximations and numerical dispersion errors are
further reduced in the modified incompressibility algorithm by
\citet{Dietrich1997}.

Horizontal resolution is the same in both the longitudinal
($\phi$) and latitudinal  ($\lambda$) directions, with  $\Delta
\phi$= (1/8) of degree and $\Delta \lambda=\cos\lambda \Delta
\phi$ thus making square horizontal control volume boundaries.
Vertical resolution is variable, with 30 control volume layers.
The thickness of control volumes in the top layer is 10.3 m and
they are smoothly increased up to the deepest bottom control
volume face at 2750 m. Thus ETOP05 bathymetry is truncated at 2750
m depth and it is not filtered or smoothed.

Horizontal viscosity and diffusivity values are constant and equal
to 10 m$^2$ s$^{-1}$. For the vertical viscosity and diffusivity,
a  formulation based on the Richardson number developed in
\citet{Pacanowski81} is used, with background values set at
near-molecular values ($10^{-6}$ and $2 \times 10^{-7}$
m$^2$s$^{-1}$ respectively). We use monthly mean wind stress
reanalyzed from 10 m wind output from ECMWF, as chosen for the
Mediterranean Sea Models Evaluation Experiment
\citep{Beckers2002}. The heat and the freshwater fluxes used to
force the model are model-determined from monthly climatological
SST and SSS as described in \citet{Dietrich2004}. The only open
boundary is the Strait of Gibraltar, where inflow conditions are
set similar to observations and outflow is model-determined by
upwind. Everywhere else, free-slip lateral boundary conditions are
used. All bottom dissipation is represented by a conventional
nonlinear bottom drag with a coefficient of 0.002. Lateral and
bottom boundaries are thermally insulating. The model is
initialized at a state of rest with the annual mean temperature
and salinity fields taken from the climatological data. The
spin-up phase of integration is carried out for 16 years. Each
year is considered to have 12 months 30 days length each (i.e. 360
days). The climatological forcings we use are adequate to identify
the mechanisms and processes occurring under typical or average
circumstances. Under this approach, high frequency motions are
weak in our model and a daily sampling is adequate. The impact on
transport of disturbances containing high frequencies, such as
storms or wind bursts, is not the focus of the present Paper and
would need specific modelling beyond climatological forcing.

We focus on velocity fields obtained at the second layer which has
its center at a depth of 15.93 m. This is representative of the
surface circulation and is not as directly driven by wind as the
top layer. We have recorded velocities, temperatures and
salinities in this model layer for five years. Dynamical systems
approaches have already been applied to this data set, in
particular Lyapunov techniques to quantify mixing strength
\citep{dOvidio2004}, and the ``leaking'' approach
\citep{Schneider2005} to quantify escape and residence times in
several areas. Here we concentrate in the Northwestern region and
apply the methods of lobe dynamics to characterize transport
processes in the North Balearic Front area.

\begin{figure}
  \centering
 \psfig{width=.9\textwidth, figure=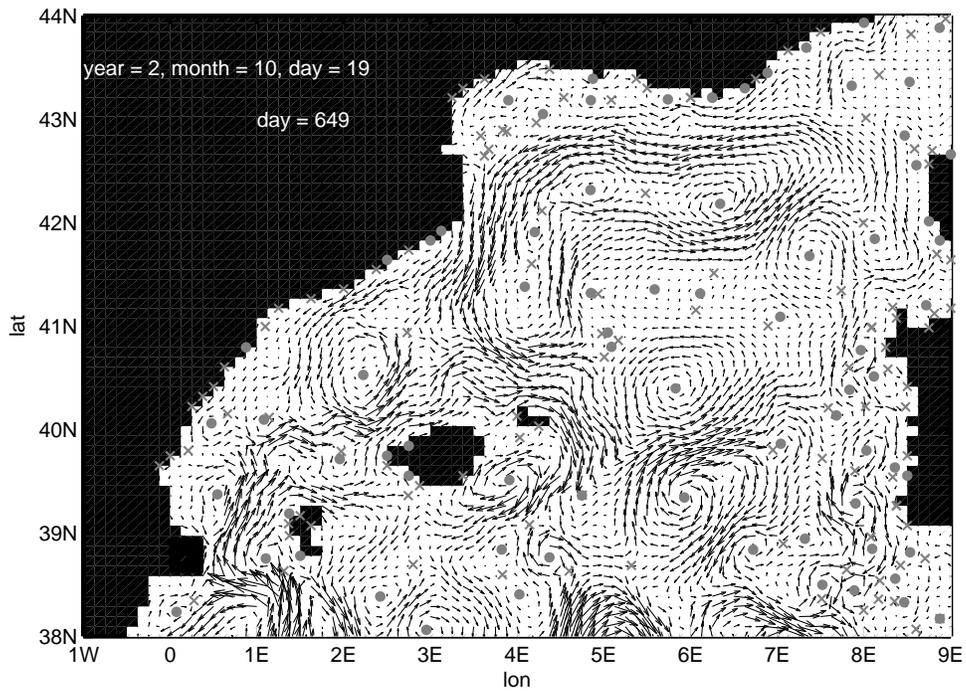 }
\caption{Velocity field at 15.93 m depth in the Northwestern
Mediterranean at simulation day 649 (October). The Westernmost
coast is the Spanish one, the islands are the Balearics, and the
Easternmost coasts are portions of Corsica and Sardinia. Circles
indicate elliptic ISPs and crosses indicate saddle-type ISPs.}
\label{velocityfield}
\end{figure}

Figure \ref{velocityfield} shows an example of the output of the
model for the velocity field in the selected layer of the Western
Mediterranean Sea at day 649 (the 19th day of the tenth month
--October-- of the second year). Two well known currents, the
Northern Current flowing southwards close to the Spanish coast and
the Balearic Current associated with the North Balearic Front and
flowing northeastwards North of the Balearics, are observed
although significantly deformed by the presence of eddies. The
Figure shows also the ISPs of the velocity field: circles for the
elliptic and crosses for the saddle-type ones.

The velocity field has small vertical components, so that this is
not strictly a two dimensional flow. With vertical velocities of
the order of $10^{-5}$ $m/s$, particles in the second model layer
require about 13 days to traverse the layer. But during that time
this vertical velocity is not constant. Averaging over the
relevant time scales we find effective velocities of 0.1-0.7
m/day, depending on location and season
\citep{dOvidio2004,Schneider2005}, and thus residence times in the
second layer are between two weeks and several months. As a rule
of thumb we can consider that trajectories preserve two
dimensionality during time intervals of about 20 days. Since most
of our trajectory integrations will be restricted to time
intervals below that duration, they can be considered two
dimensional to a good approximation.

By inspection of the temperature and salinity model outputs we
identify a four month interval starting on October of the second
simulation year (i.e. Autumn and early Winter) as an example in
which gradients are strong most of the time and thus the North
Balearic Front well defined (see Fig.
\ref{fig:SalinityManifolds}). We select this interval of time as
our study case for which dynamic structures will be calculated.

\section{Computation of Trajectories and Manifolds in Mediterranean Data Sets}
\label{sec:computation}

The equations of motion that describe the horizontal evolution of
particle trajectories in our velocity field are
\begin{eqnarray}
\frac{d \phi}{d t}&=&\frac{u(\phi,\lambda,t)}{R {\rm cos}(\lambda)}\\
\frac{d \lambda}{d t}&=&\frac{v(\phi,\lambda,t)}{R}
\end{eqnarray}
where $u$ and $v$ represent the eastwards and northwards
components of the surface velocity field coming from the
simulations described in the previous section, $R$ is the radius
of the Earth (6400 km in our computations), $\phi$ is longitude
and $\lambda$ latitude. Particle trajectories must be integrated
in equations (1)-(2) and since information is provided just in a
discrete space-time grid, a first issue to deal with is that of
interpolation of discrete data sets. A recent paper by \citet{msw}
compares different interpolation techniques in tracking particle
trajectories. Bicubic spatial interpolation in space \citep{nr}
and third order Lagrange polynomials in time are shown to provide
a computationally efficient and accurate method. We use this
technique in our computations. However we notice that bicubic
spatial interpolation in space as discussed in \citet{nr} requires
an equally spaced grid. Our data input is expressed in spherical
coordinates, and the grid is not uniformly spaced in the latitude
coordinate. In order to interpolate in an uniformly spaced grid,
we transform our coordinate system ($\lambda, \phi$) to a new
coordinate system with coordinates ($\mu, \phi$). The latitude
$\lambda$ is related to the new coordinate $\mu$
\begin{equation}
\mu=   {\rm ln}|{\rm sec}\lambda+ {\rm tan}\lambda| \label{mul}
\end{equation}
Our velocity field is now on a uniform grid in the ($\mu, \phi$)
coordinates. The equations of motion in the old variables are
transformed to a new expression in the new variables,
\begin{eqnarray}
\frac{d \phi}{d t}&=&\frac{u(\phi,\mu,t)}{R {\rm cos}(\lambda(\mu))} \label{sd1}\\
\frac{d \mu}{d t}&=&\frac{v(\phi,\mu,t)}{R {\rm
cos}(\lambda(\mu))} \label{sd2}
\end{eqnarray}

where $\lambda(\mu)$ is obtained by inverting Eq. (\ref{mul}),
i.e.
\begin{equation}
\lambda=\frac{\pi}{2}-2 \,{\rm atan}(e^{-\mu}) \ . \label{lmu}
\end{equation}

Once trajectories are integrated from these equations, for
presentation purposes one can convert $\mu$ values back to
latitudes $\lambda$ just by using (\ref{lmu}).

Distinguished hyperbolic trajectories and their unstable and
stable manifolds are the main dynamical systems objects that we
use to describe and quantify transport. Our velocity data set
lasts only for a finite time, it is highly aperiodic in time, and
turbulent. Computation of hyperbolic trajectories for this kind of
flow is discussed in \citet{jsw} and in \citet{idw}, where
algorithms for their computation are developed. A novel technique
to compute  stable and unstable manifolds of hyperbolic
trajectories in aperiodic flows is developed in \citet{mswi}. In
\citet{msw2} these hyperbolic trajectory and manifold computation
algorithms are combined into a unified algorithm and successfully
applied to a turbulent wind driven, quasigeostrophic double-gyre.
We will now apply these same algorithms to compute hyperbolic
trajectories and their stable and unstable manifolds that we will
use to describe and quantify transport associated with the North
Balearic front.

\section{Lagrangian structures and transport in the Balearic Sea}
\label{sec:results}

\begin{figure}[htb!]
\begin{center}
\includegraphics[width=5in]{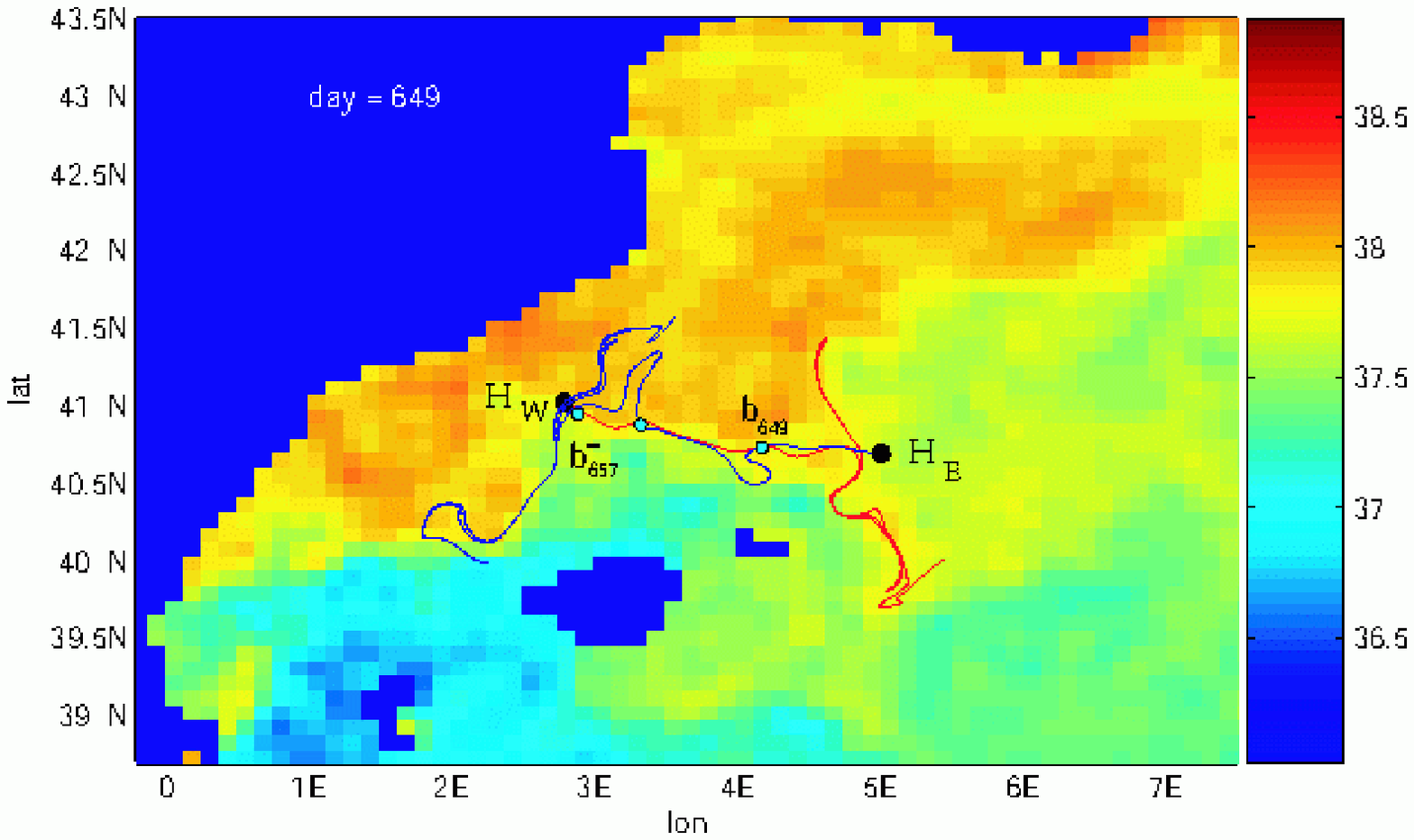}
\includegraphics[width=5in]{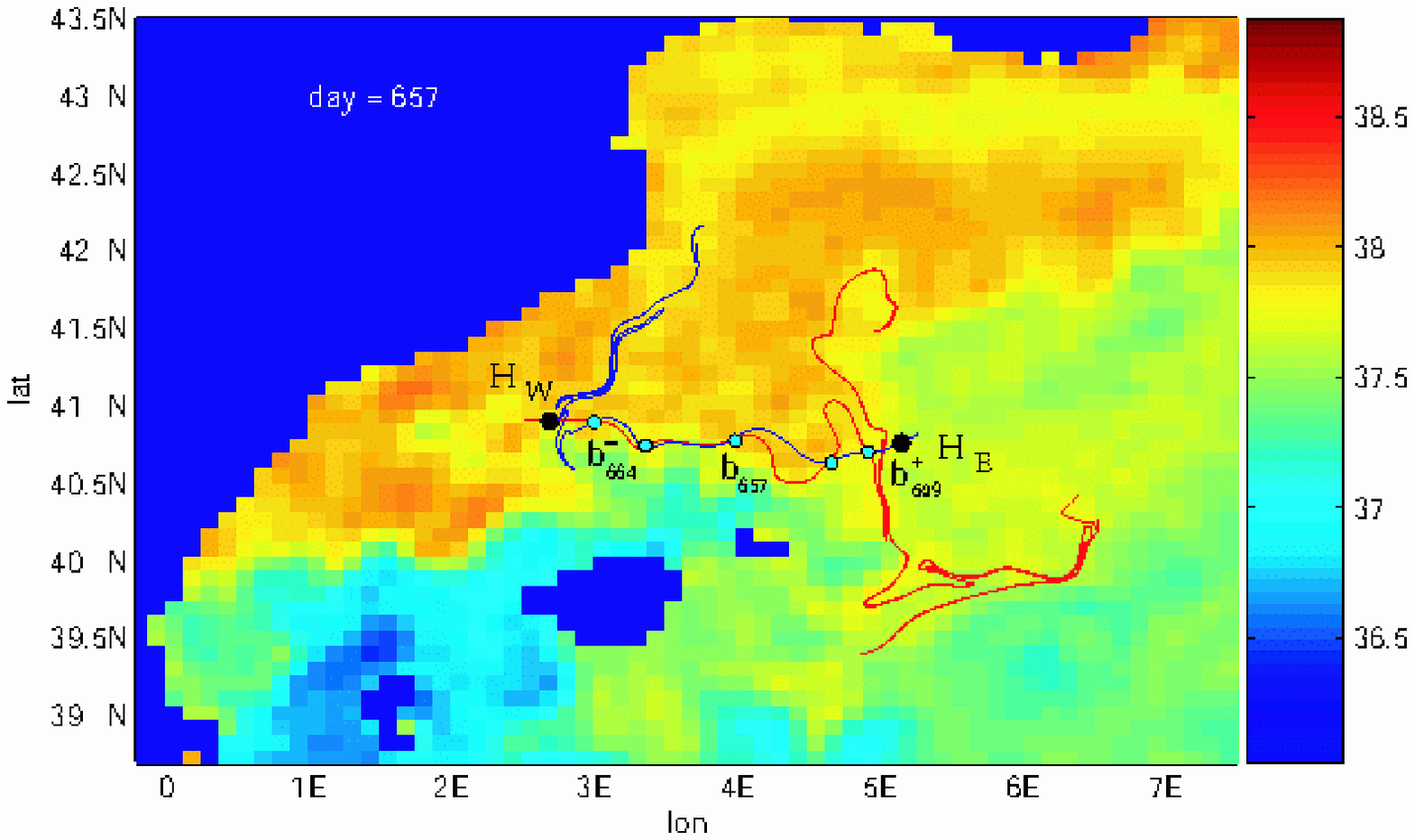}
\end{center}
\caption{Salinity front and manifolds at days 649 and 657.
Salinity (in psu) is color coded as indicated by the color bar.
The black circles denote the DHT in the West ($H_W$) and the DHT
in the East ($H_E$). These are the DHTs whose unstable (red) and
stable (blue) manifolds, respectively, are used to construct the
Lagrangian boundary of the front. The blue dots are boundary
intersection points, as described in the text.}
\label{fig:SalinityManifolds}
\end{figure}

We focus on the region North of the Balearic Islands, the Balearic
Sea. The main oceanic structures known to be present there are the
Balearic current and the associated North Balearic Front
\citep{LopezGarcia1994,Millot1999}. This last feature is known to
represent a transition zone between saltier and fresher waters in
the Western Mediterranean. The salinity fields obtained from our
computer simulation display significant salinity gradients (and
also temperature gradients in Winter--in Summer the surface layer
is heated in a rather homogeneous way) in the area (see Fig.
\ref{fig:SalinityManifolds}). Our aim here is to interpret the
presence of the gradients and the front in terms of a
semipermanent ``Lagrangian barrier'' across which little transport
occurs. This construction would also reveal the routes along which
this transport happens. Topological changes in that picture,
associated with the crossing by eddies and that may be interpreted
as the breakdown of the front, are also observed during the
simulation.

Gradients are rather well defined during Autumn of the second
simulation year and early Winter of the third one. During this
period we find a long interval (from day 649 to day 731) in which
a Lagrangian structure constructed using stable and unstable
manifolds of DHTs remains persistent and acts as a partial barrier
to transport. Its location is well correlated with the salinity
gradients, so that it can be interpreted as a Lagrangian
identification of the North Balearic Front. The weak transport
across the structure can be described in terms of lobe dynamics.
The situation resembles the one in \citet{cw} in which
quasigeostrophic dynamics has been used to model a double gyre
situation and the central jet between the two gyres in that
problem  plays a similar role as that of the Balearic current in
our problem. Lobe dynamics was there successfully applied to
quantify transport across the jet, occurring by the so called
turnstile mechanism. However in the more realistic data set
analyzed here we need to generalize some ideas used there. For
example, in  \citet{cw} DHTs stay on one dimensional boundaries
for all time (as boundaries are invariant), however in our
scenario the relevant current does not start nor end on clear one
dimensional boundaries. This introduces some ambiguity in the
identification of the relevant DHTs (and of the saddle-type ISPs
used as starting positions in the iterative algorithm of
\citep{msw2} that we use to determine the DHTs) from which to
compute the manifolds that will define the Lagrangian barrier.
Most of the time, however, pieces of manifolds computed from
different DHTs in the same area rapidly converge towards each
other, thus indicating that the location of the dominant
hyperbolic curve is not a property of the particular choice of
DHTs, but a property of the flow. Changes in the topology of the
flow cause this convergence property to be lost. This happens at
the end of the interval of time chosen in our study case and will
be commented on below.

\subsection{Cross frontal transport: the turnstile mechanism}

The geometric objects used to characterize transport by lobe
dynamics methods are constructed by the rules discussed in
\citep{maw, cw, samwig}. Here we describe in some detail these
procedures in a way that is particular to our flow situation.
Figs.  \ref{fig:SalinityManifolds} - \ref{fig:lobes} illustrate
the construction at days 649 and 657. Fig.
\ref{fig:SalinityManifolds} contains all the dynamic structures
superimposed on a salinity field, whereas for clarity only the
'boundary' and the 'lobes' are displayed in Figs.
\ref{fig:boundaries}, and \ref{fig:lobes}, respectively.

\begin{enumerate}

\item Two DHTs should be identified, one in the western part of
the front to be characterized and another in the eastern, that
persists close to their initial positions during the whole time
interval of interest (denoted by $[t_0,t_N]$). Since our algorithm
to locate DHTs uses saddle ISP positions as first guesses, figures
displaying ISPs such as Fig. \ref{velocityfield} are used to
estimate these positions and the temporal persistence of the ISPs.
The positions calculated for the selected DHTs are plotted in
Figs. \ref{fig:SalinityManifolds} - \ref{fig:break} as black dots
and labelled as $H_W$ (the western one) and $H_E$ (the eastern
one).

\item As the mean current flows eastwards we proceed as in
\citet{cw} and compute the unstable manifold of the western DHT
and the stable manifold of the one in the East. They are the red
and blue lines in Fig. \ref{fig:SalinityManifolds}, respectively.
For clarity in the presentation, of the two branches of each
manifold (one at each side of the DHT from which they emanate) we
display in Fig. \ref{fig:SalinityManifolds} only the one pointing
in the direction of the other DHT. Along both pieces of manifolds,
in the region between the two DHTs, the dominant direction of
motion is from west to east. As characteristic in unsteady flows,
both manifolds intersect repeatedly. Some of the intersection
points are marked with cyan dots. It is also typical that the
unstable manifold (red) of a DHT ($H_W$ in this case) emerges from
it relatively straight, whereas it fluctuates widely when
approaching the vicinity of the opposite DHT. In the same way, the
stable manifold (blue) of $H_E$ joins it smoothly, but displays
characteristic oscillations when close to the western DHT. The
stable and unstable manifolds displayed in Fig.
\ref{fig:SalinityManifolds} a) have been computed after backwards
and forwards time integration of a small segment in the direction
of the stable and unstable linear subspaces  in the neighborhood
of the DHT for time periods of 14 and 19 days, respectively.
Integrations for longer time periods provide longer manifolds.
However due to the restrictions of the two dimensional
approximation, a time period beyond 20 days is not completely
trustworthy. In practice this means that the pieces of the
manifold that are far from the DHT may  deviate from the true
manifold as those pieces are the ones obtained with longer time
integrations. For instance, the unstable manifold in Fig.
\ref{fig:SalinityManifolds} b) has been computed for a time
integration of 27 days. This means that unstable structure in the
far East is not completely reliable. However the piece of manifold
governing the turnstile mechanism, which is closer to the DHT, is
obtained with numerical integrations below the validity limit of
20 days and predictions obtained from them are correctly
approached in the two dimensional approximation.

\item Between the beginning and the end of the chosen time
interval, we choose a sequence of times at which to analyze the
manifold positions and compute objects relevant for Lagrangian
transport. The sequence of ``observation times'' is denoted by
$t_0 < t_1 < t_2 < \ldots < t_{N-1} < t_N$. We note that the $t_i$
do {\em not} need to be equally spaced. In order to illustrate the
construction of boundaries and the turnstile mechanism for
crossing the boundaries we need only two times, and for this
purpose we will choose to show days 649 and 657.

\item At each of the selected times $t_i$ , a ``boundary'' is
  constructed by choosing a finite piece of the unstable manifold of
  the western DHT, $[H_W,b_{t_i}]$, and a finite piece of the stable
  manifold of the eastern DHT, $[b_{t_i},H_E]$, so that they intersect
  in precisely one point, $b_{t_i}$, which is called a {\em boundary
    intersection point}. The points $\{b_{t_i}\}$, in addition to
  satisfying an ordering constraint specified below, should be
  selected in such a way that a boundary relatively straight, i.e.
  free of the violent oscillations displayed by each of the manifolds
  when approaching the opposite DHT, is obtained. Since the boundary
  is pinned at the points $H_W$ and $H_E$ we obtain a sequence of
  boundaries that fluctuate in time but remain approximately in the
  same place. Figure \ref{fig:SalinityManifolds} shows the selection of
  the boundary intersection points $b_{649}$ and $b_{657}$ at times
  649 and 657, respectively. For clarity, Fig. \ref{fig:boundaries}
  displays just the resulting boundaries.

  \begin{figure}[htb!]
\begin{center}
\includegraphics[width=5in]{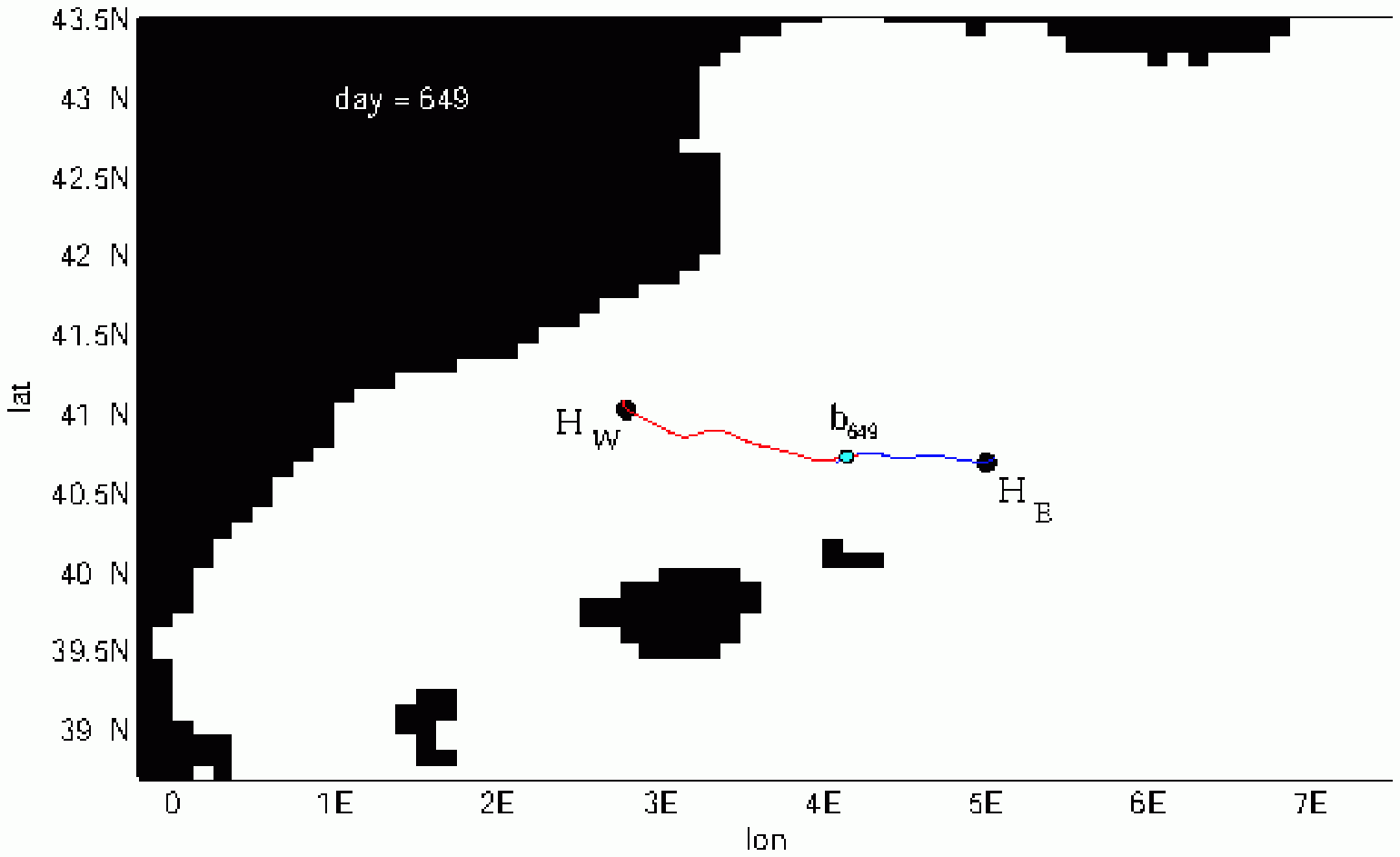}
\includegraphics[width=5in]{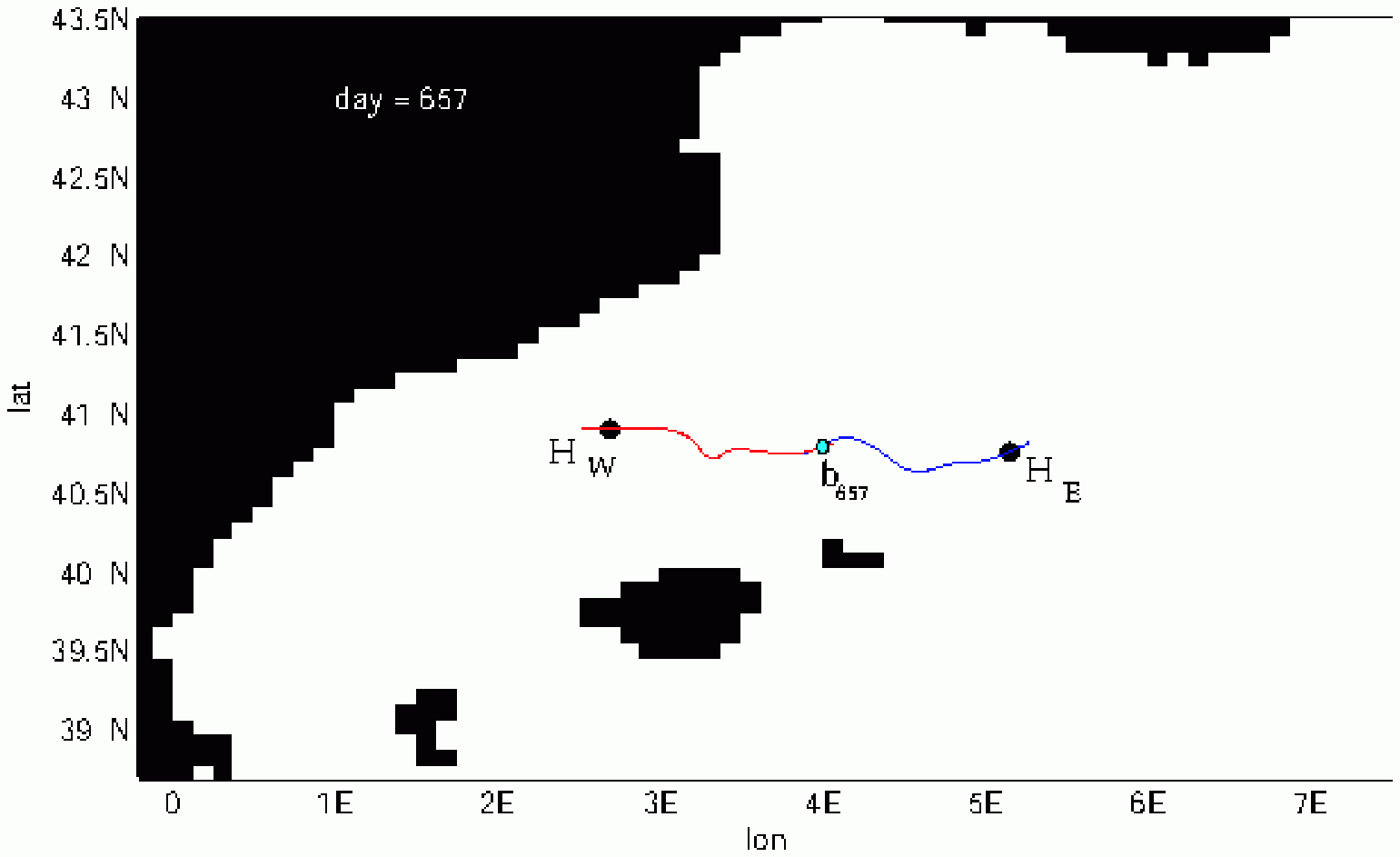}
\end{center}
\caption{Boundaries at days 649 and 657 constructed from a (finite
length) segment of the unstable manifold of $H_W$ and a (finite
length) segment of the stable manifold of $H_E$. The {boundary
intersection points} are denoted by $b_{649}$ and $b_{657}$,
respectively.} \label{fig:boundaries}
\end{figure}

  Since the boundary is made of material lines, no fluid can cross it by horizontal advection
  processes, except at the observation times $t_1, t_2, ... , t_N$ at
  which the boundary is redefined. At these times, the only way fluid
  at one side of the boundary can be transferred to the other side is
  by the turnstile mechanism described and quantified below. If this
  transport amount is small (as it will be shown to be the case), the
  boundary can be characterized as a ``barrier'', and gradients will
  be maintained across it. As seen in
  Fig.~\ref{fig:SalinityManifolds}, the position of the boundary is
  well correlated with the position of the salinity front thus
  confirming that the dynamical systems techniques developed here are
  useful to identify the North Balearic front in terms of Lagrangian
  objects.

\item Construct turnstiles at $t_i$. At time $t_i$ we consider the
  point, denoted by $b_{t_{i+1}}^-$, that will evolve into the
  boundary intersection point $b_{t_{i+1}}$ (clearly, we cannot do
  this at $t_N$). Since the stable and unstable manifolds are
  invariant, this point is also on both the stable and unstable
  manifolds. In the same way, $b_{t_{i-1}}^+$ is the location at time
  $t_i$ of the boundary intersection point that was located at
  $b_{t_{i-1}}$ at the previous time $t_{i-1}$. The additional
  constraint that needs to be imposed when choosing the sequence
  $\{b_{t_i}\}$ is that $b_{t_{i+1}}^-$ results ``upstream'' (i.e.
  closer to $H_W$ along its unstable manifold, or further from $H_E$
  along its stable manifold) from $b_{t_i}$. This introduces a
  restriction on the choice of $b_{t_{i+1}}$ once $b_{t_i}$ is chosen.
  Since ordering of points along manifolds is preserved by time evolution, it
  turns out that $b_{t_{i-1}}^+$ would be ``downstream'' from
  $b_{t_i}$. Figures~\ref{fig:SalinityManifolds}, \ref{fig:boundaries}
  and \ref{fig:lobes} display some of these intersection points,
  showing that our selection satisfies the ordering constraints. The
  segments of stable and unstable manifolds between $b_{t_{i}}$ and
  $b_{t_{i+1}}^-$ (and between $b_{t_{i}}$ and $b_{t_{i-1}}^+$) trap
  regions of fluid, and these regions of fluid defined in this way are referred to as {\em lobes}. Some of them
  can be seen in Fig. \ref{fig:SalinityManifolds}, and more clearly
  in Fig. \ref{fig:lobes}.

\item Consider the evolution of the turnstile lobes from $t_i$ to
$t_{i+1}$. In this case $b_{t_{i+1}}^-$ evolves to the boundary
intersection point, $b_{t_{i+1}}$, and the boundary intersection
point at $t_i$ evolves to a point $b_{t_i}^+$, which is also on
both the stable and unstable manifolds. Then the lobes between
$b_{t_{i+1}}$ and $b_{t_i}^+$ represent the time evolution of the
turnstile lobes from $t_i$ to $t_{i+1}$. Note that, because of the
way in which the boundary is redefined at each observation time,
and because of the different shape of the manifolds when close or
far from the DHT from which they emanate, each lobe is on opposite
sides of the boundary at the two considered times. These turnstile
lobes contain all the fluid that has crossed the boundary between
$t_i$ and $t_{i+1}$. This transport process is illustrated more
clearly in Fig. \ref{fig:lobes} where the lobes experiencing the
turnstile mechanism are plotted before and after crossing the
boundary.

\begin{figure}[htb!]
\begin{center}
\includegraphics[width=5in]{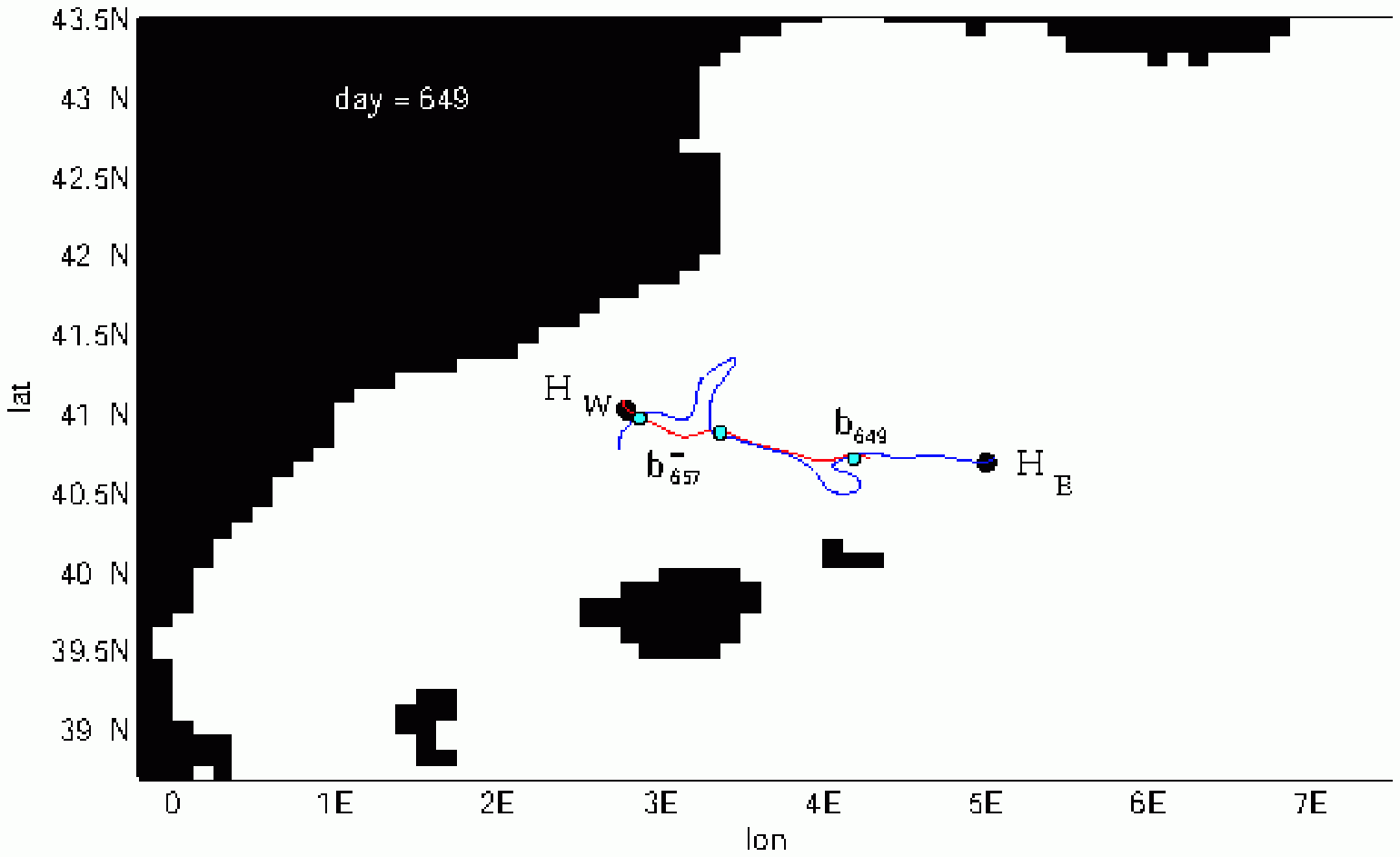}
\includegraphics[width=5in]{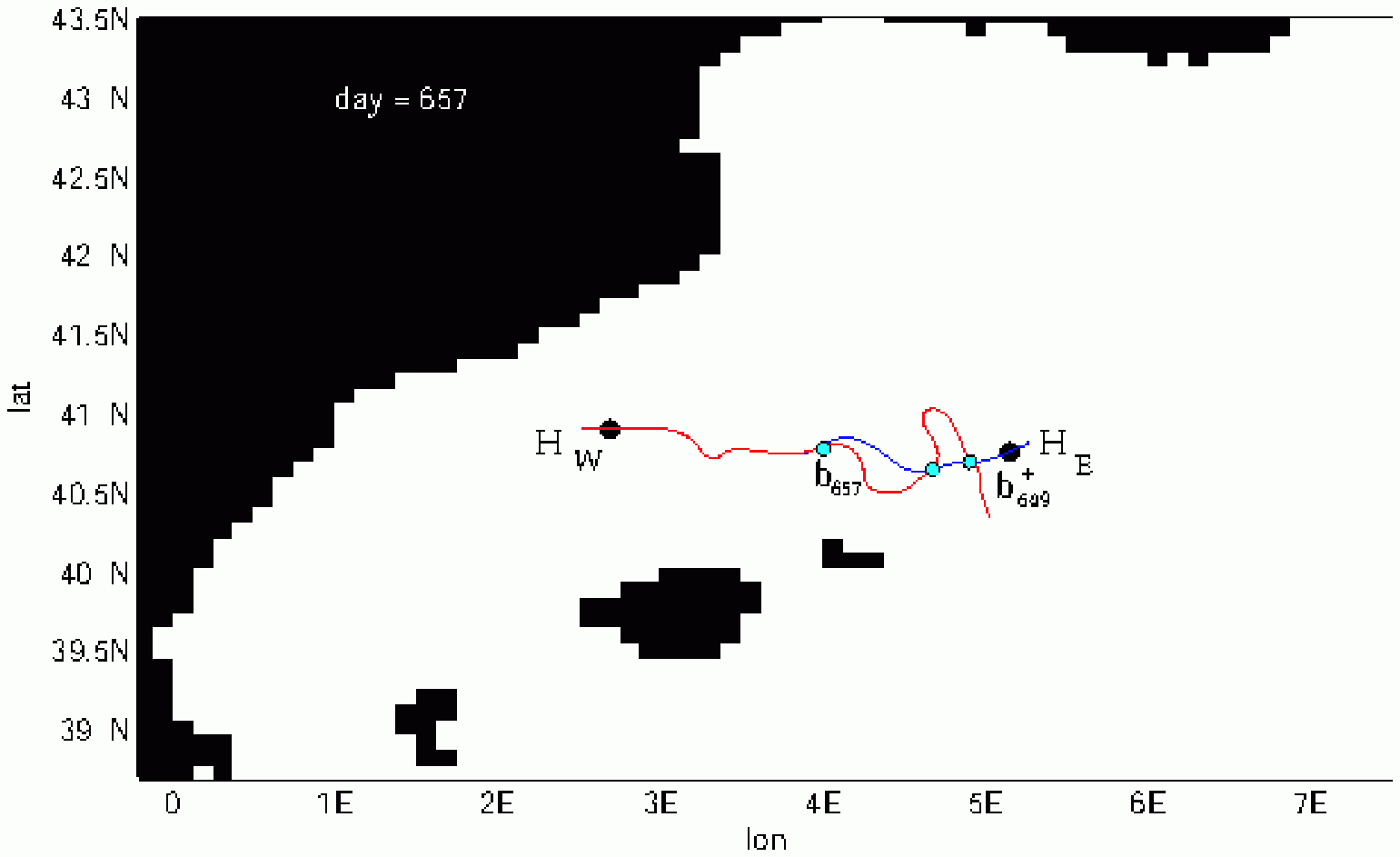}
\end{center}
\caption{Turnstile lobes from day 649. There is precisely one
intersection point between $b_{649}$ and $b^-_{657}$, which
implies that there are two lobes in the turnstile. The bottom
figure shows the evolved position of these lobes at day 657.
Comparing the two figures, one can see that the turnstile lobe to
the North of the boundary at day 649 has moved to the South of the
boundary at day 657, and the turnstile lobe to the South of the
boundary at day 649 has moved to the North of the boundary at day
657 (the location of the boundary itself can be seen in
Fig.~\ref{fig:boundaries}). } \label{fig:lobes}
\end{figure}

\end{enumerate}

The geometric construction performed at every observation time
$t_i$ as explained above allows us to calculate the amount of
transport occurring across the boundary during each time interval.

In computing the area $A(\Gamma)$ of a lobe $\Gamma$ we use the
formula

\begin{equation}
A(\Gamma)=-R^2 \int_{\partial\Gamma} \sin\lambda d\phi
\end{equation}

\noindent
 where the integration is around the closed curve which
forms the boundary of the lobe. That this formula gives the area
can be seen by considering the differential form $\omega \equiv
-R^2 sin\lambda d\phi$, calculating its differential
\citep{spivak}: $d\omega  = R^2 cos\lambda d\phi d\lambda$ which
is identical to the area element on the sphere in spherical
coordinates, and recalling Stokes theorem \citep{spivak}:

\begin{equation}
   \int_{\partial\Gamma} \omega = \int_\Gamma d\omega.
\end{equation}

For day 649, as shown in Fig.~\ref{fig:lobes}, the two turnstile
lobes have areas of $493.2 km^2$ (the lobe below the boundary, to
the east) and $716.9 km^2$ (the lobe to the west above the
boundary).  At day 657, the eastern lobe is above the new
boundary, and the western lobe is below it.  Assuming that the
divergence of the surface flow can be neglected so that the areas
are unchanged -- in numerical experiments we have never observed
more than a $3\%$ change -- we can calculate the flux across the
barrier to be $(716.9-493.2) km^2= 223.7 km^2$, in the southerly
direction.

Modified Atlantic Waters occupy the surface layers of the area
until an average depth of about $150 m$. Using as an approximation
for the mean horizontal speed of this water mass the values at the
second model layer considered here, multiplication of the $150 m$
depth by the area of the lobes gives $33.56\times 10^9 m^3$ in $8$
days, or an average flux over this interval of 0,049 Sv (1Sv=$\rm
10^6 m^3/s$). The average flux obtained from area calculations of
further turnstile lobes until the middle of November remains below
that value (the average is 0.025 Sv, always southwards). This
should be compared to the 0.75-0.5 Sv which are transported by the
Balearic current. We see that cross-boundary transport is small
during this time interval and thus the Lagrangian boundary acts as
a "barrier to transport'' permitting only small amounts of mixing
between Northern and Southern waters. It will maintain a salinity
(and thus density) front that we identify with the observed North
Balearic front. There is some indeterminacy in the definition of
the boundary that we identify as the front, arising from some
freedom in the selection of the the intersections $\{b_{t_i}\}$.
But other choices can only displace the boundary by a distance of
the order of the size of the lobes, which we see is small when not
too close to the DHTs, and in fact also of the order of the width
of the transition region in salinity distributions such as the one
in Fig. \ref{fig:SalinityManifolds}, i.e. the width of the front.

\subsection{Spatio-temporal structure of cross frontal transport}

\begin{figure}[htb]
\begin{minipage}{\pwidth}
  \centering
  \subfigure[Evolution of the turnstile lobes at  day 674 backwards in time to day 667,
  or, in other words, the location on day 667 of the lobes that are turnstile lobes at day 674.
  The notation $b_{681}^{--}$ denotes the evolution of the boundary intersection point at day
  681 {\em backwards} two observation times (to day 674 and to the current time, day 667).  ]
  {\psfig{width=\fwidth, figure=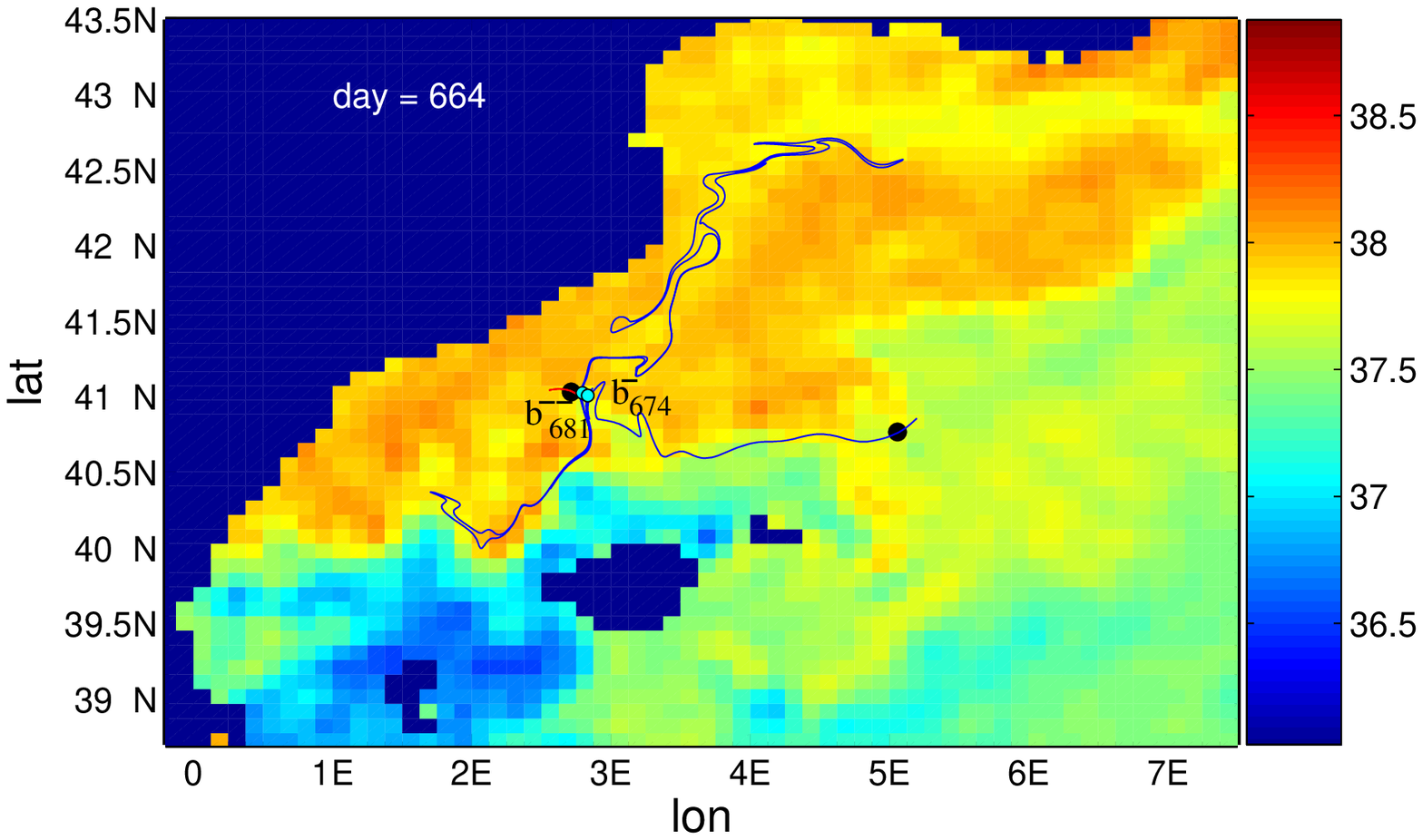}} \\
  \subfigure[Turnstile lobes  at day 674.]
  {\psfig{width=\fwidth, figure=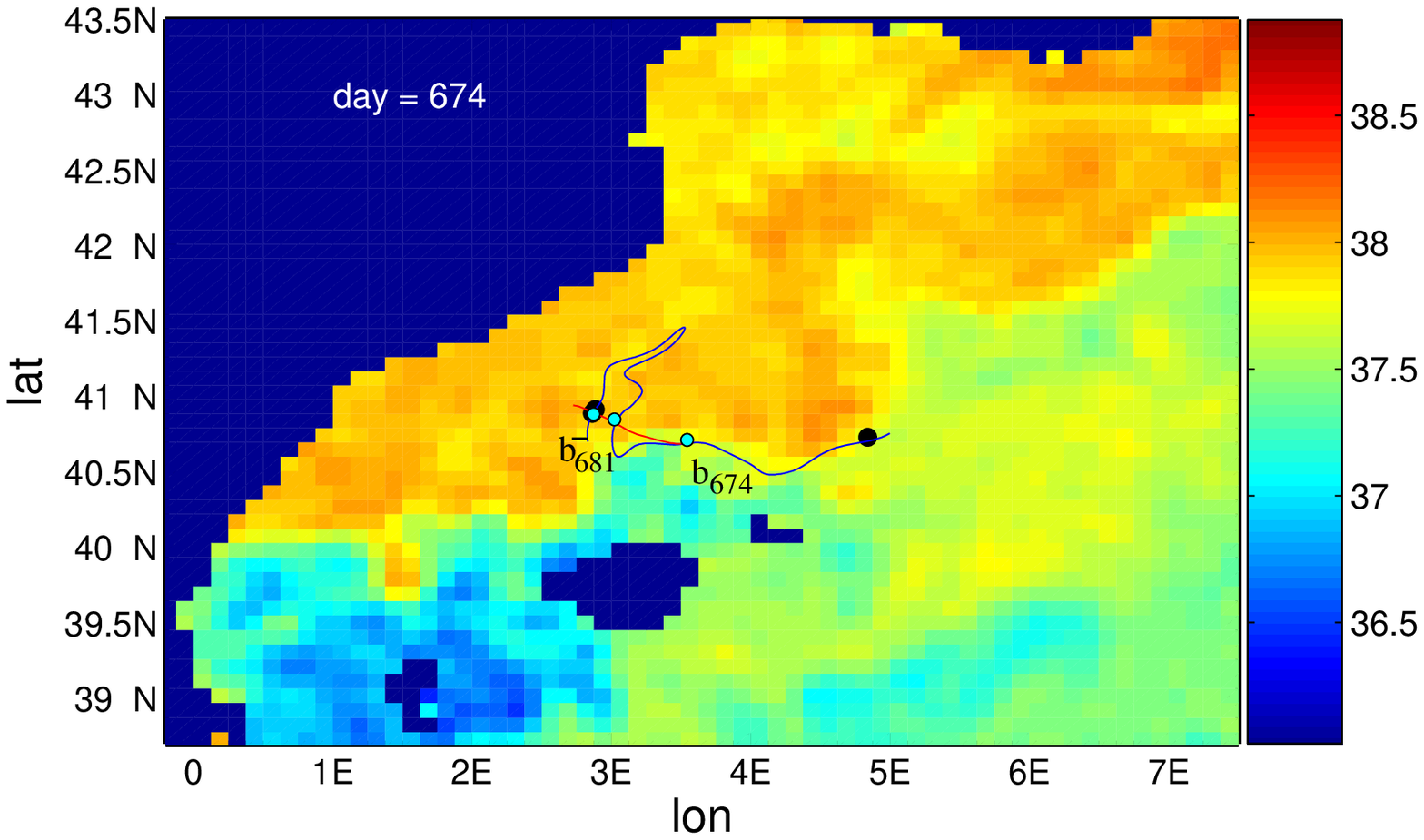}} \\
\end{minipage}
\begin{minipage}{\pwidth}
  \centering
  \subfigure[The turnstile lobes evolved to day 681.]
  {\psfig{width=\fwidth, figure=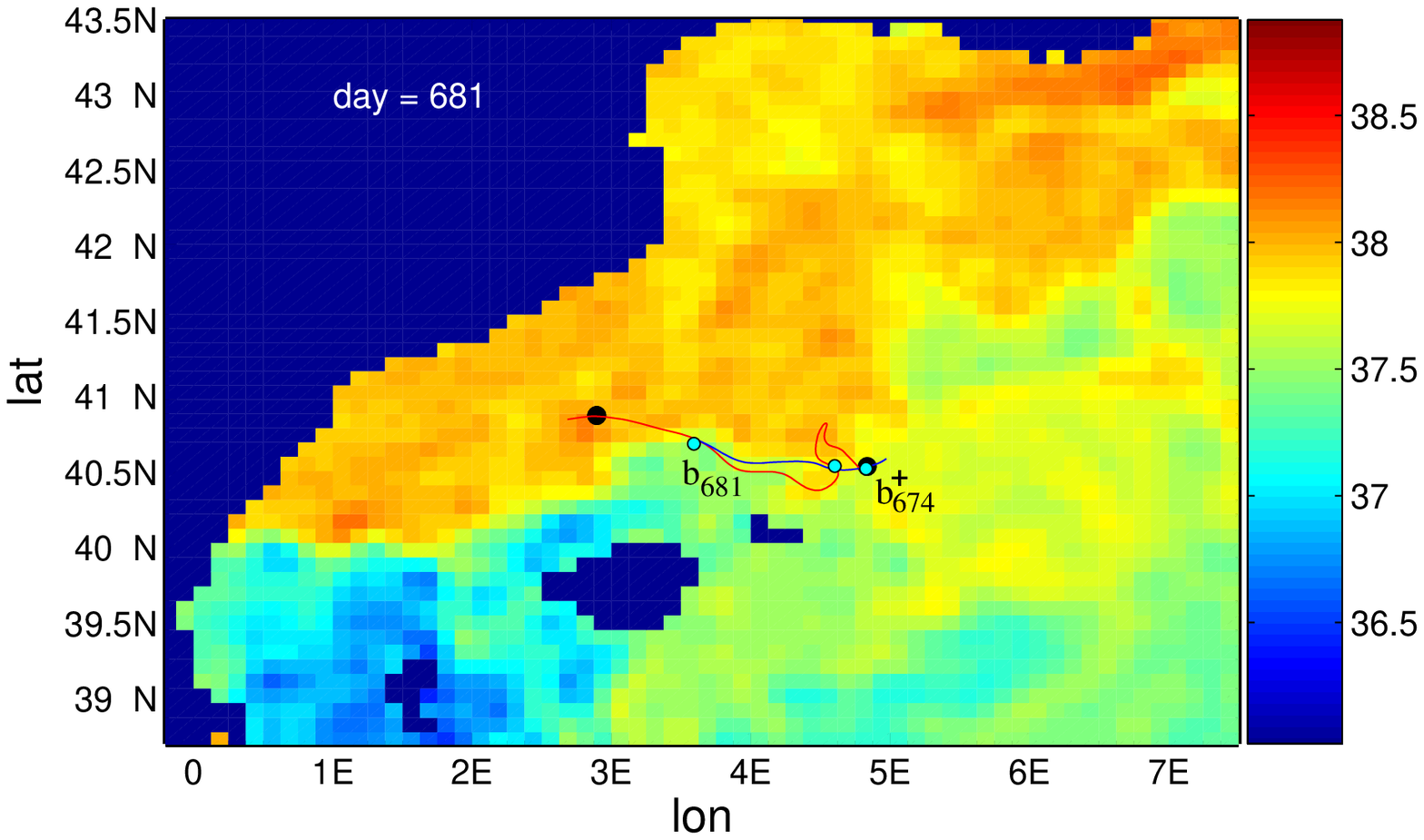}} \\
  \subfigure[The turnstile lobes evolved to day 688. The notation $b_{674}^{++}$ denotes the evolution of the boundary intersection
  point at day 674 {\em forwards} two observation times (to day 681 and to the
current time, day 688).]
  {\psfig{width=\fwidth, figure=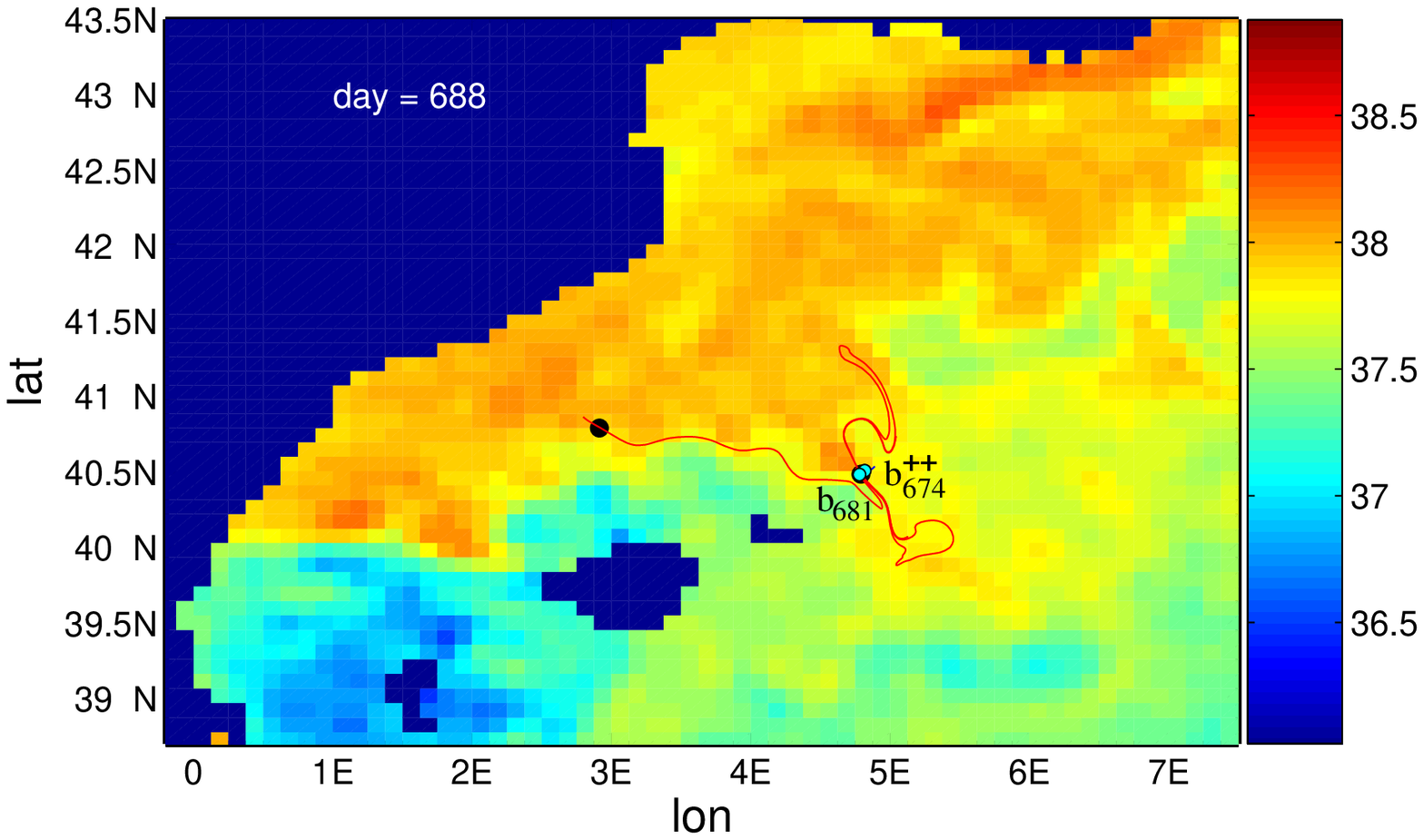}} \\
\end{minipage}
\caption{The spatio-temporal structure of cross frontal transport
described by lobe dynamics. Salinity (in psu) is color coded as
indicated by the color bar.} \label{fig:LobeHistory}
\end{figure}

Since the only way in which our constructed Lagrangian boundaries
can be crossed is via the turnstile mechanism, the earlier and
later location of the turnstile lobes reveals the dominant routes
along which the weak cross-front transport occurs. Figure
\ref{fig:LobeHistory} shows the time evolution of a pair of
turnstile lobes, one initially above and the other initially below
the boundary, as they evolve in time. The crossing of the boundary
by the turnstile mechanism occurs between days 676 and 681, and
the remaining panels in the figure show the position of these
lobes at some earlier and later times. The sequence illustrate
that lobes move essentially along the boundary except when close
to $H_E$, where they are ejected as filaments transverse to the
front (the one initially in the south ejected towards the north
and reciprocally for the one started in the north) and when close
to $H_W$, where they have also the shape of transverse filaments
and they become entrained into the boundary region. Fig.
\ref{fig:LobeHistory} also illustrates how lobes transport water
of different salinity (coded in colors) and how the above routes
for lobe motion and shape correlate with the salinity distribution
in the area. Note that the length of the manifolds and the whole
process depicted in Fig. \ref{fig:LobeHistory} lasts only 21 days,
so that the plotted manifolds remain approximately horizontal,
with only small corrections from the vertical flow.

\subsection{An Eddy-front interaction: disruption of the
Lagrangian boundary}

\begin{figure}[htb!]
\begin{minipage}{\pwidth}
  \centering
  \subfigure[Day 733. The unstable manifold of $H_W$ does not appear to intersect the stable manifold of $H_E'$. The unstable manifold is strongly
  influenced by the circulation associated with the elliptic ISP immediately to  the lower right of $H_E'$.]
  {\psfig{width=\fwidth, figure=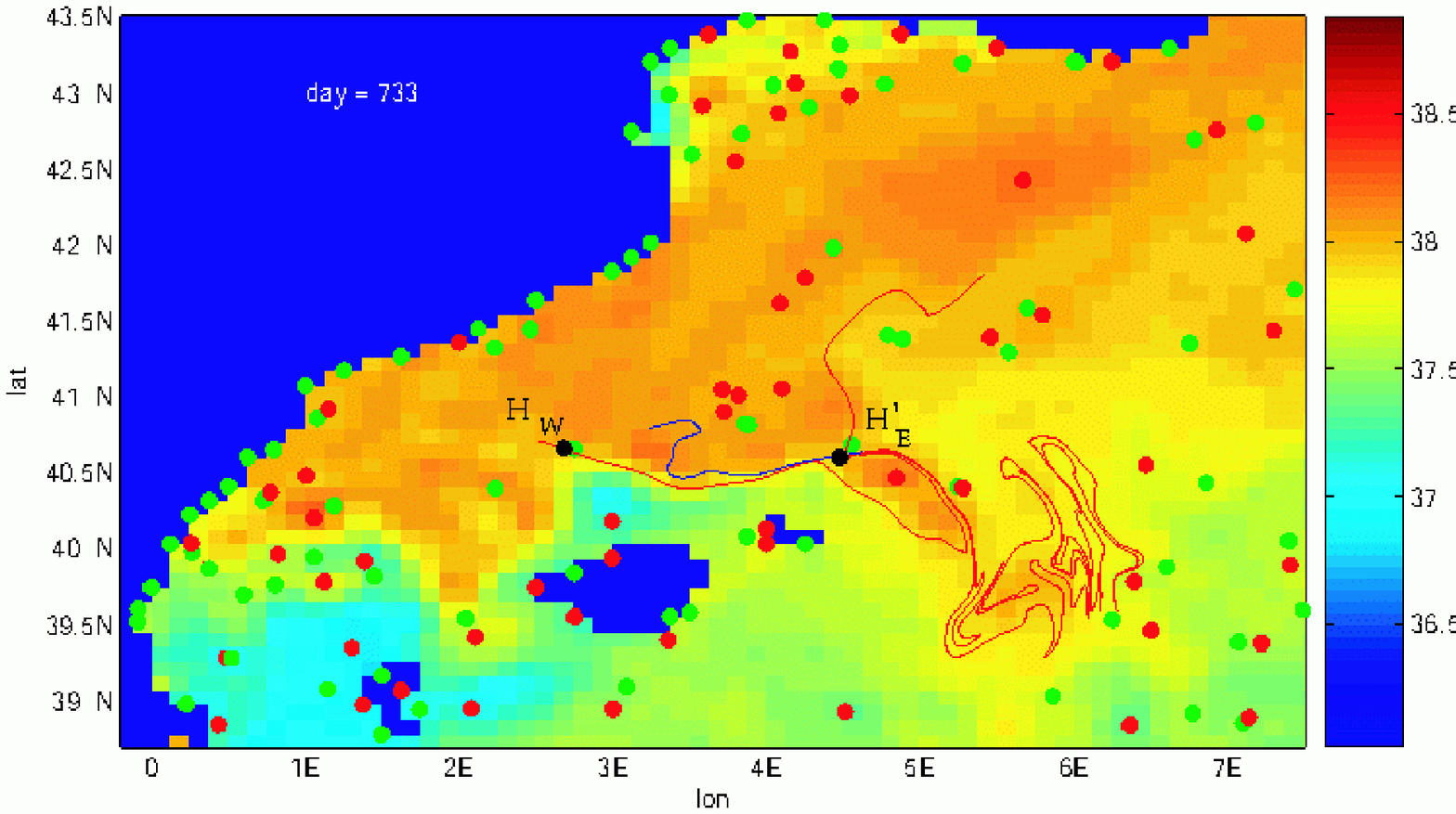}}
  \subfigure[Day 736.  As the elliptic ISP moves further away from $H_E'$ its influence carries the unstable manifold with it.]
  {\psfig{width=\fwidth, figure=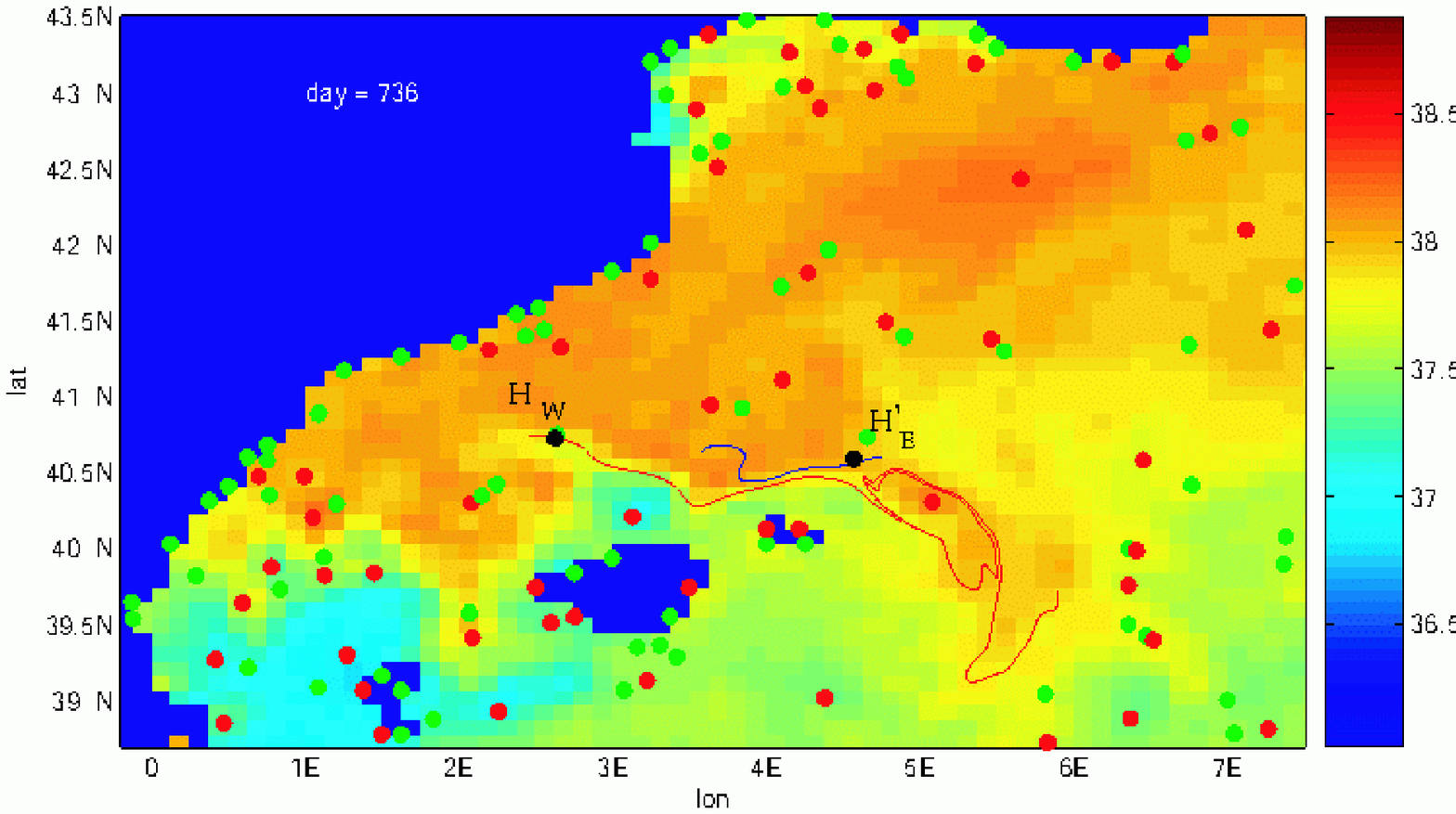}} \\
  \end{minipage}
\begin{minipage}{\pwidth}
  \subfigure[Day 742. The earlier Lagrangian barrier now appears completely destroyed as a result of the influence of the eddy. ]
  {\psfig{width=\fwidth, figure=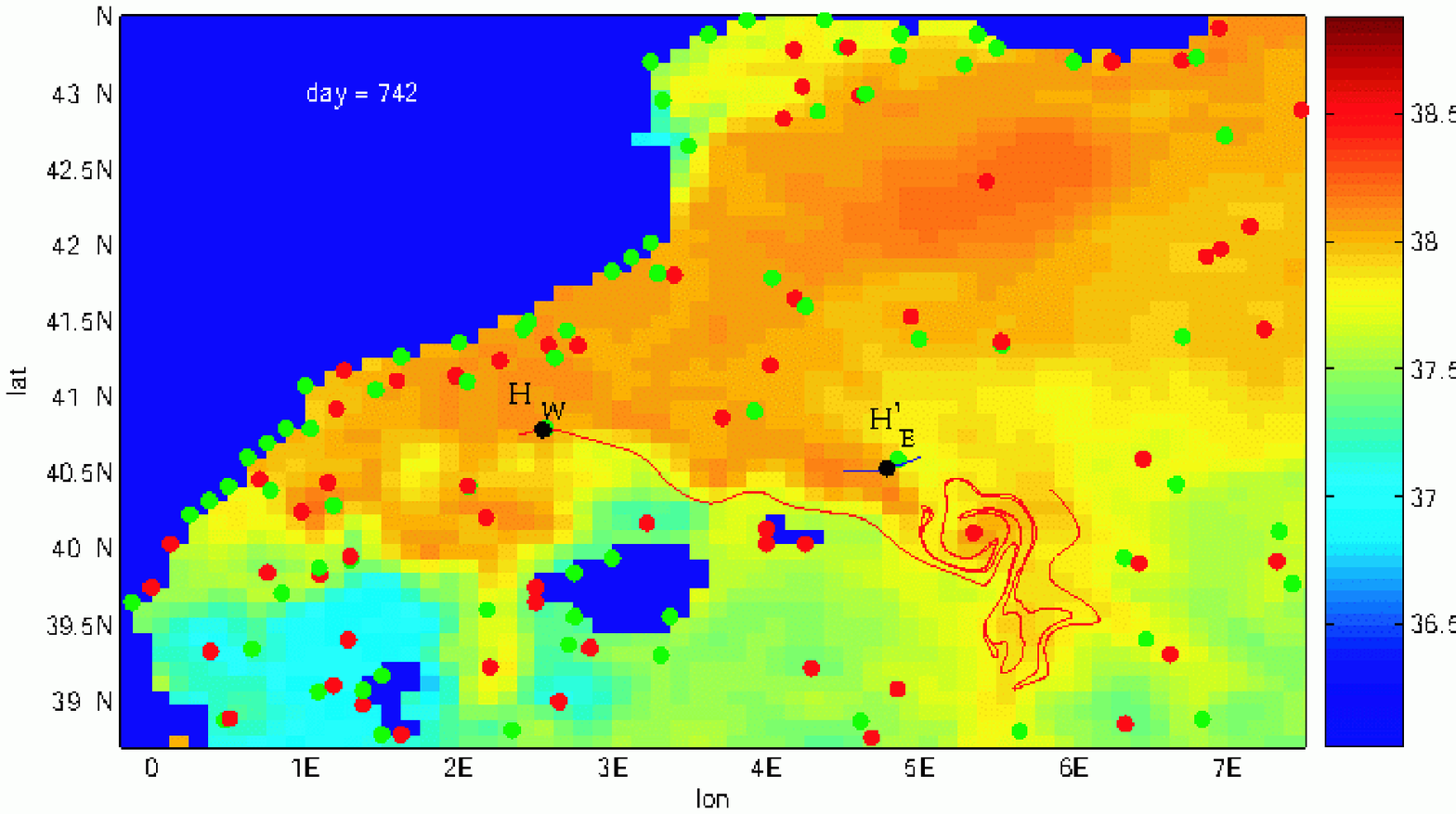}}
  \end{minipage}
  \caption{The mechanism of the disruption of the Lagrangian barrier by an eddy. Salinity (in psu) is coded in colors as indicated by the color bar.
  Red dots are ISPs of elliptic character. Green dots are saddle ISPs. Also shown as black dots are the DHTs $H_W$ and $H'_E$.}
  \label{fig:break}
\end{figure}

Not all flow configurations allow the geometric construction
identifying the Lagrangian boundary and the associated turnstile
lobes to be performed. It may happen that no pair of DHTs persist
in a given area long enough to support the mechanism, or their
manifolds can fail to intersect. It may also happen that manifolds
started at relatively close DHTs do not converge to the same curve
but remain significatively distinct, so that a unique well defined
boundary can not be properly identified. In such cases, the
turnstile transport mechanism is not the most relevant one. At the
end of the simulation interval analyzed here we observe a change
in the topology that signals the end of the predominance of the
turnstile mechanism in a process that can be interpreted as the
breakdown of the front by the interaction with an eddy. As a first
symptom, calculations of turnstile lobe areas reveal an increase
in cross-front transport starting at day 674 (middle of November).
The average transport between days 674 and 700 is of 0.303 Sv
(southwards), still smaller than the Balearic current transport
but significantly larger than the average cross-frontal transport
during the previous month (0.025 Sv, also southwards). At day 711,
stable manifolds emerging from $H_E$ and from another rather close
DHT cease to converge into each other, signaling the end of a
situation with an essentially unique well defined boundary. Later,
at day 731 our algorithm is unable to find the location of $H_E$
starting from ISPs in the area. This probably means that $H_E$ has
moved away from the area under study. The black dot in Fig.
\ref{fig:break} is another DHT found in the region. But its stable
manifold (i.e., the finite length of manifold that we are able to
compute) does not intersect the unstable manifold from $H_W$, thus
revealing that is in fact a DHT different from $H_E$, and that it
can not support the turnstile mechanism. The time evolution of the
unstable manifold from $H_W$ suggests that the reason from the
change in behavior is the breakdown of the Lagrangian barrier by
the crossing of an eddy, identified by the rolling up of the
manifold around an elliptic ISP (Fig. \ref{fig:break}). Note that
even in this situation the manifold position is well correlated
with the salinity distribution, thus indicating that still
Lagrangian structures are relevant. But the transport mechanism is
clearly different from the turnstile described above, being more
appropriately described as water transport inside an eddy.

\section{Conclusions}
\label{sec:conclusions}

  In this work we have applied in a systematic way some tools
developed in the context of dynamical systems theory and known
generically as `lobe dynamics'. The computer generated surface
velocity field studied here is more complex and less regular that
other velocity fields previously considered in this context, but
we have found that one of the main mechanisms of transport by lobe
motion, the turnstile, is still at work. The methodology includes
the construction of a `barrier' across which to compute transport,
and in our application to the Northwestern Mediterranean dynamics
it has been identified with one of the main oceanographic
structures present here, the North Balearic Front. Transport
across it proceeds in the form of filaments that are entrained
into the front close to a DHT `upstream', and released also in the
form of filaments close to another DHT located `downstream'. The
ejection of these filaments at that location can explain recent
observations of waters saltier than expected just east of the
island of Menorca \citep{Emelianov2006}. The identification of the
DHTs and the calculation of their locations is by itself an
important subject, since they organize the flow in the area and,
because of this and of the sensitivity of the trajectories in
their neighborhood, they are candidates for launch locations in
efficiently designed drifter release experiments
\citep{ptkj,Molcard2006}.

Despite the success of the approach described here, much work
remains to be done in order to develop dynamical systems
techniques into a collection of systematic tools for  analyzing
general oceanographic data. A classification and understanding of
the different topological regimes leading to qualitatively
different modes of transport and the transitions among them,
similar to the existing ones for steady and time-periodic flows,
would be desirable for the cases of turbulent aperiodic flows. A
characteristic of the dynamical systems approach is that it
provides an unusually high detailed description of the
spatio-temporal structure of Lagrangian transport. Therefore it
may well turn out to be the optimal tool for analyzing data from
 ocean models with much higher spatio-temporal resolution
(e.g., higher frequency atmospheric forcing and more resolved
spatial scales) which capture more physical processes. This would
enable us to better define, for example, the process of the
destruction of a barrier. In addition, consideration of the impact
of non-Lagrangian processes such as diffusion, of vertical
motions, and of strong localized perturbations beyond
climatological forcing such as storms, would be needed to have a
more complete vision of transport phenomena and mechanisms.

\section*{Acknowledgements}
A.M.M. acknowledges the  MCyT (Spanish Government) for a Ram\'on y
Cajal Research Fellowship and financial support from MEC (Spanish
Government) reference MTM2004-00797 and the Royal Society-CSIC
cooperation agreement reference B2003GB03. E.H.-G. acknowledges
financial support from MEC and FEDER through Project CONOCE2
(FIS2004-00953). We also acknowledge M. Emelianov for
communicating us results from a recent cruise before publication.
D.S. and S. W. acknowledge financial support from ONR Grant
No.~N00014-01-1-0769.

\end{document}